\documentclass{article}
\usepackage{dsfont}
\usepackage{dcase2021_techrep,amsmath,graphicx,url,times,booktabs, tabularx}
\usepackage{xcolor}
\usepackage{subcaption}
\usepackage{hyperref}
\usepackage{enumitem}
\usepackage{algorithm}
\usepackage{algorithmic}
\usepackage{tablefootnote}
\usepackage{colortbl}

\definecolor{aqua}{rgb}{0.0, 1.0, 1.0}
\definecolor{awesome}{rgb}{1.0, 0.13, 0.32}

\title{A Lottery Ticket Hypothesis Framework for Low-Complexity Device-Robust Neural Acoustic Scene Classification}

\name{Hao Yen$^{1}$, Chao-Han Huck Yang$^{1}$,
      Hu Hu$^{1}$,
       Sabato Marco Siniscalchi$^{1,2}$, Qing Wang$^{3}$, Yuyang Wang$^{3}$,
      }
\secondlinename{
      Xianjun Xia$^{4}$,
      Yuanjun Zhao$^{4}$,
      Yuzhong Wu$^{4}$,
      Yannan Wang$^{4}$,
      Jun Du$^{3}$,
      Chin-Hui Lee$^{1}$
      }

\address{$^1$School of Electrical and Computer Engineering, Georgia Institute of Technology, GA, USA \\
$^2$Computer Engineering School, University of Enna Kore, Italy\\
$^3$University of Science and Technology of China, HeFei, China \\
$^4$Tencent Media Lab, Tencent Corporation, China
}

\begin{document}

\ninept
\maketitle

\begin{sloppy}

\begin{abstract}
We propose a novel neural model compression strategy combining data augmentation, knowledge transfer, pruning, and quantization for device-robust acoustic scene classification (ASC). Specifically, we tackle the ASC task in a low-resource environment leveraging a recently proposed advanced neural network pruning mechanism, namely Lottery Ticket Hypothesis (LTH),  to find a sub-network neural model associated with a small amount non-zero model parameters. The effectiveness of LTH for low-complexity acoustic modeling is assessed by investigating various data augmentation and compression schemes,  and we report an efficient joint framework for low-complexity multi-device ASC, called \emph{Acoustic Lottery}. Acoustic Lottery could largely compress an ASC model and attain a superior performance (validation accuracy of 79.4\% and Log loss of 0.64) compared to its not compressed seed model. All results reported in this work are based on a joint effort of four groups, namely GT-USTC-UKE-Tencent, aiming to address the ``Low-Complexity Acoustic Scene Classification (ASC) with Multiple Devices'' in the DCASE 2021  Challenge Task 1a.

\end{abstract}


\section{Introduction}
\label{sec:intro}
Acoustic scene classification (ASC) aims to recognize a set of given environment classes (e.g., airport and urban park) from real-worlds sound examples. Analysis and learning to predict acoustic scene sounds are important topics associated with various mobile and on-device intelligent applications~\cite{yang2021unsupervised}. The Detection and Classification of Acoustic Scenes and Events (DCASE) challenges~\cite{dcase2016, dcase2018, dcase2017, dcase2020} provide a comprehensive evaluation platform and benchmark data to encourage and boost sound scene research communities. DCASE 2021 Task 1a~\cite{martin2021low} focuses on developing low-complexity acoustic modeling (AM) solutions for predicting sounds recorded from multiple devices (e.g., electret binaural microphones, smartphones, and action cameras). The goal is to design a device-robust ASC system preserving generalization power over audios recorded by different devices, and highlighting the importance of low-complexity requirements. %

From previous DCASE challenges, we observed that several competitive ASC systems~\cite{hu2020devicerobust, shim2021attentive, heittola2020acoustic} benefited from large-scale convolutional neural models combined with several data augmentation schemes, but whether we can attain the generalization power of those complex models with a low-complexity architecture is the research goal to be addressed in DCASE 2021 challenge. To this end, we focus on addressing two basic questions: (i) Are some well-performed device-robust ASC models overparameterized? (ii) Can we take advantage of some overparameterized models to design a low-complexity ASC framework on multi-device data? 

\vspace{-3mm}
\begin{figure}[th!]
  \centering
  \includegraphics[width=0.6\linewidth]{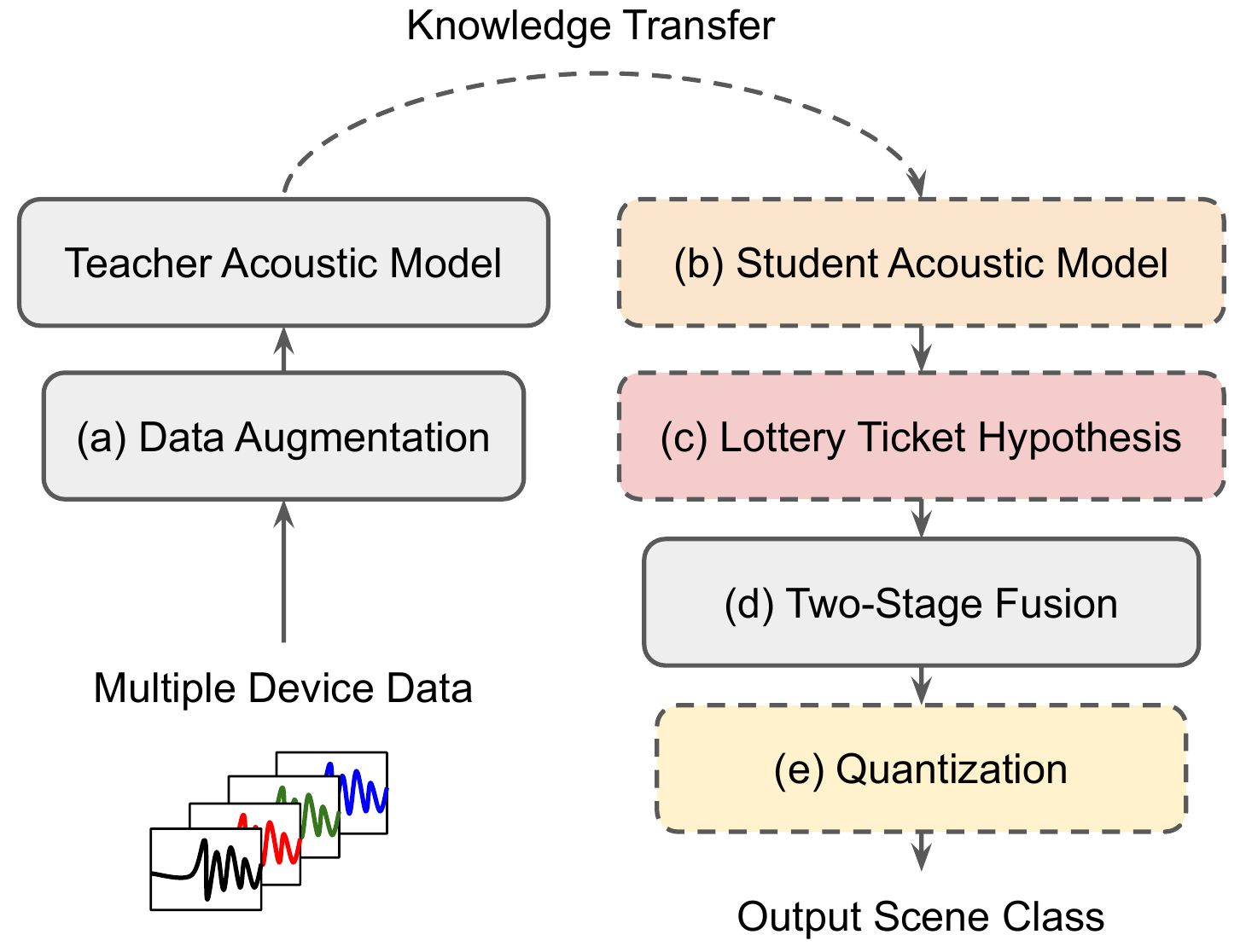}
  \caption{The proposed \textbf{Acoustic Lottery} (AL) framework.}
  \label{fig:system}
  \vspace{-3mm}
\end{figure}

In the quest for addressing the above questions, we deployed a novel framework, namely ``Acoustic Lottery,'' for DCASE 2021 Task 1a, which will be described in the following sections. As shown in Figure~\ref{fig:system}, our Acoustic Lottery system consists of (a) a data augmentation process to improve model generalization, (b) a teach-student learning mechanism to transfer knowledge, from a large teacher model to a small student model, (c) a Lottery Ticket Hypothesis~\cite{frankle2018lottery} based pruning method to deliver a low-complexity model, (d) a two-stage fusion technique \cite{hu2020two} to improve model prediction, and finally (e) a quantization block to deploy a final model owing less than 128 KB non-zero parameters, which is the requirement of Task 1a. A detailed presentation of each block in Figure \ref{fig:system}  is discussed in the following sections.
\vspace{-2mm}
\section{Low-Complexity Acoustic Modeling Framework}
\label{sec:sys}

\subsection{Data Augmentation Strategy}
\label{sec:dataaug}
In some previous works~\cite{hu2020devicerobust, hu2020two, Chen2019, Koutini2019}, data augmentation strategies played a key role to attain competitive log-loss results on the DCASE 2020 validation set~\cite{heittola2020acoustic}. With the goal of deploying a seed model with good generalization capabilities to deal with the multiple device acoustic condition, the first module (Figure.~\ref{fig:system} (a)) of our DCASE 2021 system thereby builds upon data augmentation methods investigated in \cite{hu2020devicerobust}. To be specific, we use i) Mixup \cite{mixup}, ii) Spectrum augmentation \cite{spec-aug}, iii) Spectrum correction \cite{spec-corr}, iv) Pitch shift, v) Speed change, vi) Random noise, and vii) Mix audios. The reverberation data described in \cite{hu2020devicerobust} is not used in our experiments.

\subsection{Teacher-Student Learning (TSL)}
\label{sec:ts}
Teacher-Student Learning (TSL), also named as Knowledge Distillation (KD), is a widely investigated approach for model compression \cite{hinton2015distilling, li2014learning}. Specifically, it transfers knowledge from a large and complex deep model (teacher model) to a smaller one (student model). The main idea is to establish a framework that makes the student directly mimicking the final prediction of teacher.
Formally, the soften outputs of a network can be computed by $p=softmax(\frac{\alpha}{\tau})$, where $\alpha$ is the vector of logits (pre-softmax activations) and $\tau$ is a temperature parameter to control the smoothness \cite{hinton2015distilling}. Accordingly, the distillation loss for soft logits can be written as the Kullback-Leibler divergence between the teacher and student soften outputs. In this work, we followed the approaches in \cite{hu2020devicerobust} to build a large two-stage ASC system, serving as the teacher model. Then a teacher-student learning method is used to distill knowledge to a low-complexity student model, as shown in Figure~\ref{fig:system} (b).

\subsection{Lottery Ticket Hypothesis Pruning}

Next we have investigated advanced pruning techniques to further reduce non-zero model parameters of the student. Although neural network pruning methods often negatively affect both model prediction performance and generalization power, a recent study, referred to as Lottery Ticket Hypothesis~\cite{frankle2018lottery} (LTH), showed a quite surprising phenomenon, namely pruned neural networks (sub-networks) could be trained  attaining a performance that was equal to or better than the not pruned original model if the not pruned parameters were set to the same initial random weights used for the non-pruned model. Interestingly, LTH-based low-complexity neural models had proven competitive prediction performance on several image classification tasks~\cite{frankle2018lottery} and recently have been supported with some theoretical findings~\cite{malach2020proving} related to overparameterization.

\begin{algorithm}[tb!]
  \caption{LTH for Device-Robust Acoustic Modeling}
  \label{alg:lth}
\begin{algorithmic}
  \STATE {\bfseries 1.} \textbf{Input}: a model, $G_0$; augmented sound data, $D$. \\
  \STATE {\bfseries 2.} \textbf{Randomly Initialize Weights} ($\Theta_0$).  \\
  \STATE {\bfseries 3.} \textbf{Initialize Model:} $G_0$($\Theta_0$) $\rightarrow$ $G_1$ \\
  \STATE {\bfseries 4.} \textbf{For} $t = 1, ..., T$: $\#$ \emph{Pruning Searching Iterations} \\
  \STATE {\bfseries 5.} \hspace{3mm} \textbf{For} $e = 1, ..., E$: $\#$ \emph{Gradient Training Epochs} \\
  \STATE {\bfseries 6.} \hspace{6mm} $\Theta_e$ $\rightarrow$ $\Theta$: TSL-train $G_t$ with $D$ for its final weights ($\Theta_t$) \\
  \STATE {\bfseries 7.} \hspace{3mm} \textbf{If} t $<$ T: $\#$ \emph{LTH Pruning Strategy}\\
  \STATE {\bfseries 8.}\hspace{7mm} Mask($\Theta_t$) to get a pruned graph $G_p$ from $G_0$ \\ 
  \STATE {\bfseries 9.} \hspace{6mm} Load homologous initial weights $\Theta_p \in \Theta_0$ from $G_p$\\ 
  \STATE {\bfseries 10.} \hspace{4mm} Update target model $G_p(\Theta_p) \rightarrow G_{t+1}$
  \STATE {\bfseries 11.} \textbf{Output}: A well-trained pruned model $G_T(\Theta_T)$\\
\end{algorithmic}
\end{algorithm}

\textbf{Algorithm Design:} In \textbf{Algorithm}~\ref{alg:lth}, we detail our approach  under the Acoustic Lottery framework: In step (1), we first choose a model with its original neural architecture (e.g., Inception in our case) $G_0$ and record its initial weights parameters $\Theta_0$ in step (2). In our work, we incorporate teacher-student learning framework discussed in Section~\ref{sec:ts} with the goal of mimic prediction accuracy and generalization adapted of the teacher acoustic model - a complex model trained separately. At the end of each training phase, a pruning iteration is started if the current iteration $t$ is less than $T$. The LTH pruning searches for a low-complexity model in steps (7) through (10).

From our empirical findings in DCASE 2021 Task 1A data, we found that the proposed Acoustic Lottery only needs one or two ($T$=1 or 2 in \textbf{Algorithm}~\ref{alg:lth}) searching iteration(s) to find a good low-complexity acoustic model without a significant drop in the ASC classification accuracy compared to the high-complexity teacher model on the validation set. To select the mask function in step (8), we evaluate three major LTH strategies, namely: (i) large-final; (ii) small weights, and (iii) global small weights, which were proposed in~\cite{frankle2018lottery}. We found the small weights strategy allows us to attain better trade-off between classification accuracy and compression rate  compared to the other two mentioned methods as shown in Figure~\ref{fig:lth:acc}. Therefore, we selected ``small weights'' as pruning  strategy to be used in our final submission. Finally, a well-trained pruned student acoustic model is deployed in step (10) of \textbf{Algorithm}~\ref{alg:lth}

\begin{figure}[ht!]
\begin{subfigure}[b]{0.230\textwidth}   
\centering 
\includegraphics[width=\textwidth]{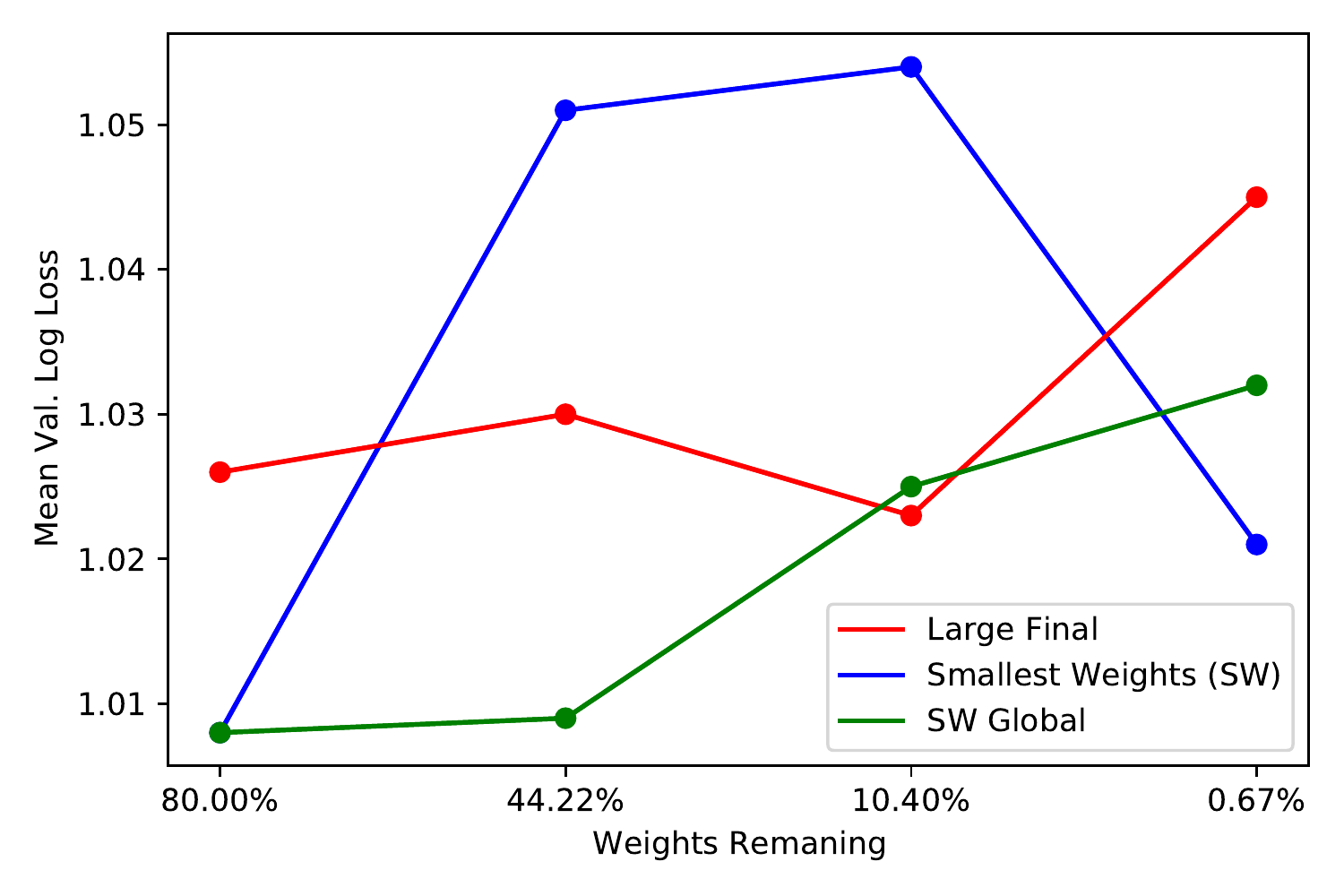}
\caption[]%
{{\small Validation Loss}}    
\label{fig:loss}
\end{subfigure}
\quad
\begin{subfigure}[b]{0.230\textwidth}   
\centering 
\includegraphics[width=\textwidth]{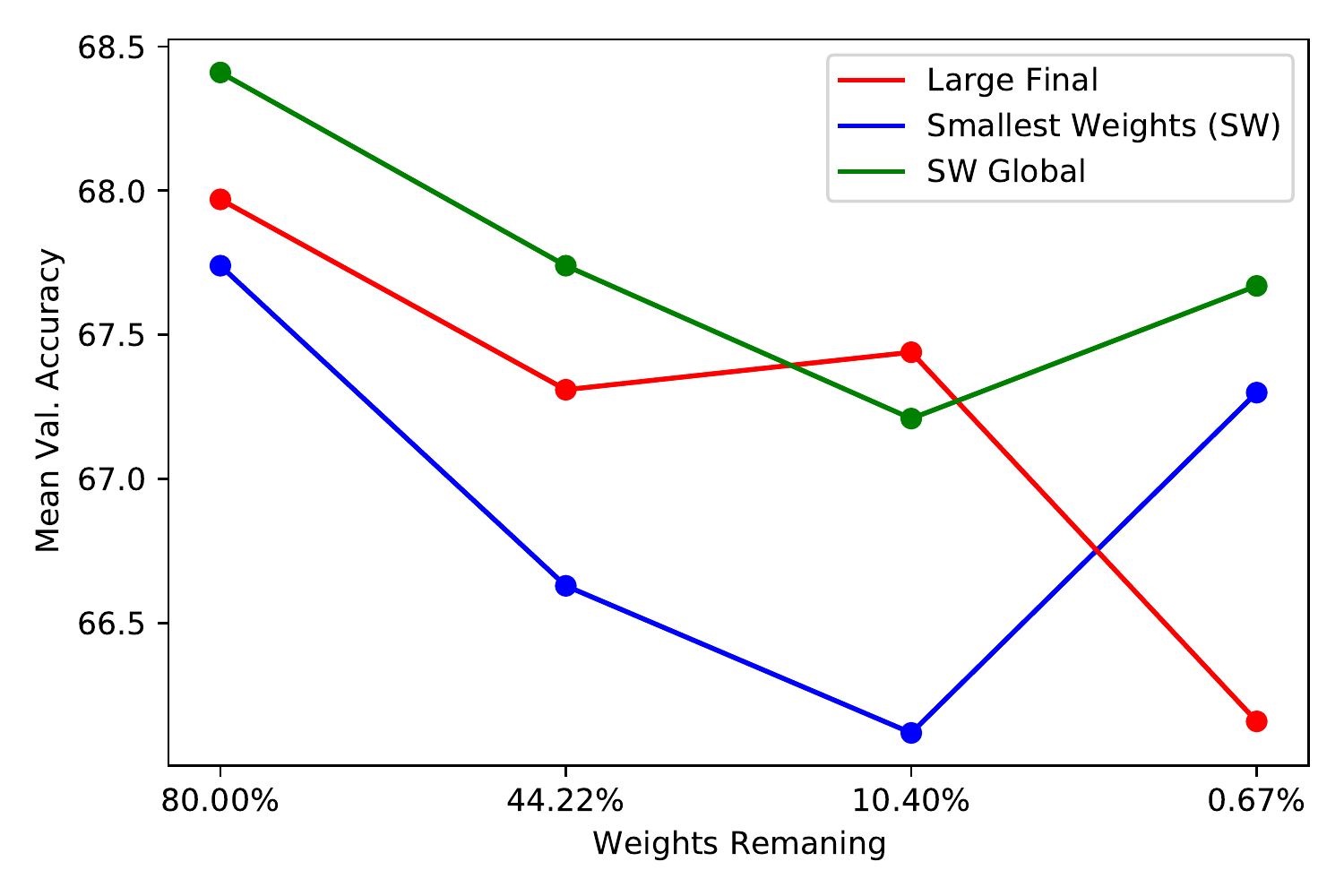}
\caption[]%
{{\small Validation Accuracy}}    
\label{fig:acc}
        \end{subfigure}
        \caption[]
        {\small We compared empirical performance of different LTH-masking strategies~\cite{zhou2019deconstructing} versus sparsity level (weights remaining). } 
        \label{fig:lth:acc}
    \end{figure}

\textbf{Visualization:} To better interpret weights distribution in an LTH-pruned neural acoustic model, we visualize a shallow inception model (excluding  convolutional layers due to their dimensional conflicts) on Index 3 in Table~\ref{tab:task1a} and its LTH-pruned results as Index 5 in in Table~\ref{tab:task1a} shown in Figure~\ref{fig:lthv}. In  Figure~\ref{fig:lthb}, we can observe that the proposed Acoustic Lottery framework can discover a well-trained model using only sparse weights with up to a $149\times$ compression rate.

\begin{figure}[ht!]
\begin{subfigure}[b]{0.215\textwidth}   
\centering 
\includegraphics[width=\textwidth]{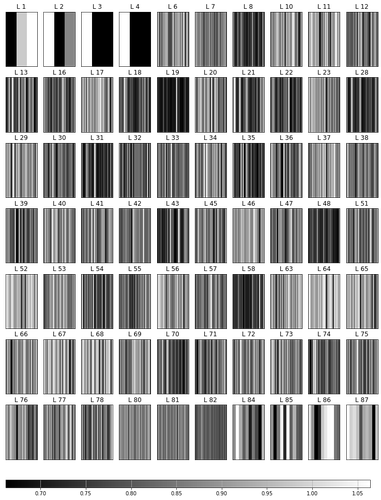}
\caption[]%
{{\small Shallow Inception (SIC)}}    
\label{fig:ltha}
\end{subfigure}
\quad
\begin{subfigure}[b]{0.215\textwidth}   
\centering 
\includegraphics[width=\textwidth]{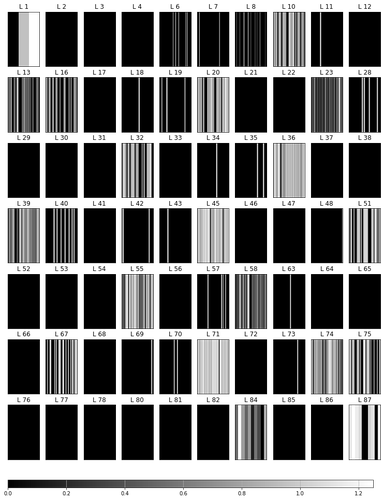}
\caption[]%
{{\small LTH-Pruned SIC}}    
\label{fig:lthb}
        \end{subfigure}
        \caption[]
        {\small Visualized of layer-wise weights distribution by the LTH approach applied to the student neural acoustic model: (a) Shallow Inception (SIC) student and (b) LTH-Pruned SIC student.} 
        \label{fig:lthv}
    \end{figure}

\vspace{-0.2cm}
\subsection{Two-Stage Fusion and Multi-Task Learning}
\label{sec:two}
\vspace{-0.1cm}
To boost ASC performance, we follow the investigation in the  two-stage ASC scheme discussed in~\cite{hu2020two}, where the relationship between the 3-class and 10-class ASC systems were exploited to boost the 10-class ASC system. This step is carried out in the module (d) in Figure~\ref{fig:system}. The key idea is that the labels of the two subtasks, 3-class and 10-class problems,  differ in the degree of abstraction and using the two labels together could be helpful.
In our setup, the 3-class classifier classifies an input scene audio into one of three broad classes: in-door, out-door, and transportation. This 3-class classification way is from our prior knowledge that scene audios can be roughly categorized into such three classes. The 10-class classifier is actually the main classifier. Each audio clip should belong to one of the three / ten classes. The final acoustic scene class is chosen by the score fusion of those two classifiers. If we let $\mathds{C}^1$ and $\mathds{C}^2$  denote the set of three broad classes, and ten classes, respectively, and let $F^1$ and $F^2$ indicate the output of the first and second classifier, respectively. The final predicted class $Class(x)$ for the input $x$ is:
\begin{equation}
    Class(x) = \underset{q, (p\in \mathds{C}^1, q\in \mathds{C}^2, p \supset q)}{\operatorname{argmax}}\ F^1_p(x) * F^2_{q}(x),\nonumber
\end{equation}
\noindent where $p \supset q$ means that $p$ can be thought of a super set of $q$. For example, transportation class is the super set for bus, tram, and metro classes. Therefore, the probability of an input audio clip to be from the public square scene is equal to the product  of the probability of out-door place, $F^1_p(x)$, and that of public square, $F^2_q(x)$.

However, the two ASC classifiers are trained separately, which means the total parameters will be doubled. In~\cite{shim2021attentive}, the authors argued that joint training of two subtasks could be even more efficient. Specifically, the 3-class classifier and the 10-class classifier can be learned in a multi-task learning (MTL)~\cite{yang2020multi} manner. The two classifiers can share some parameters, where only output layers are different. MTL is expected to perform as well as two-stage method but save parameters. We thus study that setting as an ablation module in our experimental section.

\vspace{-0.2cm}
\subsection{Quantization for Model Compression}
\label{sec:quant}
\vspace{-0.1cm}

As the main goal is to deploy a system with a size within 128 Kilobytes (KB), we further use a post-training quantization method with dynamic range quantization (DRQ), as shown in Figure~\ref{fig:system} (e). DRQ is the simplest form of post-training quantization, which statically quantizes only  weights from floating point to integer, which has 8-bits of precision. Moreover, activations are dynamically quantize based on their range to 8-bits. Leveraging upon DRQ, we thus convert our neural acoustic model from a 32-bit format to a 8-bit format, which compresses the model size to about $1/4$ of its original size as our final model.

\begin{table*}[ht!]
\centering
\caption{Experimental results on 2021 Task 1a. 'TSL' means performing teacher-student learning. 'LTH' means pruning with the Lottery Ticket Hypothesis algorithm. 'Two-stage' means using a two-stage fusion system. 'MTL' means using multi-task learning system. 'Quant' means using quantization on model parameters (float32 to float8). 'Aug' means using extra augmented data (Methods in Section~\ref{sec:dataaug}).  'System size' is according to non-zero parameters~\cite{martin2021low}. All 'Y' in the table means we used that method. Acc. indicates validation accuracy. }
\label{tab:task1a}
\begin{tabular}{c|c||c|c|c|c|c|c||c|c|c}
\toprule
\toprule
Idx. & System & TSL & LTH & Two-stage & MTL & Quant & Aug  & System size & Acc. \% & Log Loss \\ 
\midrule
(0) & Official Baseline \cite{Martnmorato2021lowcomplexity} & - & - & - & - & - & - & 90.3KB & 47.7 & 1.473 \\ 
\midrule
(1) & Two-stage FCNN \cite{hu2020two} & - & - & Y & - & - & Y & 132MB & 80.1 & 0.795 \\ 
(2) & Two-stage Ensemble \cite{hu2020two} & - & - & Y & - & - & Y & 332MB & 81.9 & 0.829 \\ 
\midrule
(3) & SIC & - & - & - & - & - & - & 503KB & 67.8 & 0.954 \\ 
(4) & SIC & Y & - & - & - & - & - & 503KB & 68.9 & 0.919 \\ 
(5) & SIC & - & Y & - & - & - & - & 296KB & 68.2 & 0.914 \\ 
(6) & SIC & - & - & Y & - & - & - & 1006 KB & 68.9 & 0.914 \\ 
(7) & SIC & - & - & - & Y & - & - & 504KB & 68.0 & 0.915 \\ 
(8) & SIC & - & - & - & - & Y & - & 126KB & 66.9 & 0.972 \\ 
(9) & SIC & Y & - & Y & - & - & - & 1006KB & 69.2 & 0.874 \\ 
(10) & SIC & Y & - & Y & - & Y & - & 252KB & 68.4 & 0.906 \\
(11) & SIC & - & Y & - & - & Y & - & 74KB & 67.7 & 0.931 \\ 
\midrule
(12) & LIC & - & - & - & - & - & - & 3434KB & 67.64 & 1.019 \\ 
(13) & LIC & Y & - & - & - & - & - & 3434KB & 67.91 & 1.002 \\
(14) & LIC & Y & Y & - & - & - & - & 2060KB & 67.64 &  \\
(15) & LIC & Y & Y & - & - & - & - & 1374KB & 68.22 &  \\
(16) & LIC & Y & Y & - & - & - & - & 687KB & 68.59 & 0.925 \\
(17) & LIC & - & - & Y & - & - & - & 7056 KB & 70.0 & 0.848 \\ 
(18) & LIC & Y & Y & Y & - & - & - & 412KB & 70.8 & 0.833 \\
(19) & LIC & Y & Y & - & - & - & Y & 687KB & 74.98 & 0.781 \\
(20) & LIC & Y & Y & Y & - & - & Y & 412KB & 78.4 & 0.654 \\
(21) & LIC & Y & Y & - & - & Y & Y & 103KB & 78.2 & 0.723 \\ 
\midrule
(22) & Ensemble of LICs & Y & Y & Y & - & Y & Y & \textbf{117KB} & 78.7 & 0.644 \\ 
(23) & Ensemble of LICs & Y & Y & Y & - & - & Y & 218KB & 79.0 & \textbf{0.637} \\ 
(24) & Ensemble of LICs and SICs & Y & Y & Y & - & - & Y & 222KB & \textbf{79.4}& 0.640 \\ 
\bottomrule
\bottomrule
\end{tabular}
\end{table*}

\section{Experimental Setup \& Results}
\label{sec:exp}
\vspace{-0.1cm}

\subsection{Feature Extraction}
We follow the same settings from DCASE 2020 Task-1a extracting acoustic features for DCASE 2021 Task-1a~\cite{martin2021low} before using the features to train low-complexity described in Section 2 and Figure~\ref{fig:system}. Log-mel filter bank (LMFB) features were used in our experiments as audio features. The input audio waveform is analyzed with a $2048$ SFFT points, a window size of $2048$ samples, and a frame shift of $1024$ samples. Thus the final input tensor size is thus $423 \times 128 \times 3$ for Task 1a. Before feeding the speech feature tensors into CNN classifier, we scaled each feature value into [0,1].

\subsection{Model Training}
\vspace{-0.1cm}
All ASC systems are evaluated on the DCASE 2020 task1a development data set \cite{dcase2020}, which consists of $\sim$14K 10-second single-channel train audio clips and $\sim$3K test audio clips recorded by 9 different devices, including real devices A, B, C, and simulated device s1-s6. Only device A, B, C, s1-s3 are in the training set; whereas, devices s4-s6 are unseen in the training phase. The greatest amount of training audio clips are recorded with device A, namely over 10K audio clips. In the test set, the number of waveforms from each device is the same. 

We use two different Inception \cite{szegedy2015going} models as our target models, namely Shallow Inception (SIC) and Large Inception (LIC). SIC has two inception blocks whereas LIC has three inception blocks and more filters. The size computed by the way recommended in \cite{Martnmorato2021lowcomplexity} of the original SIC and LIC are 503KB and 3528KB, respectively. All Inception models in this work are built with Keras based on Tensorflow2. Stochastic gradient descent (SGD) with a cosine-decay-restart learning rate scheduler is used to train all deep models. Maximum and minimum learning rates are 0.1, and 1e-5, respectively. In our final submission, all development data is used. And due to there is no validation data, we use the output of model when learning rate hits the minimum number.

\vspace{-0.2cm}
\subsection{Results on Task 1a}
In Table~\ref{tab:task1a}, we report only some of the evaluation results for low-complexity models collected on Task 1a due to space constraints.
Two inception models: (i) shallow inception model (SIC) and (ii) large inception model (LTC), are investigated under the proposed Acoustic Lottery framework. By evaluating several low-complexity strategies shown in~\ref{tab:task1a}. 
From the results, Index (0) is the official baseline, which has the size of 90.3KB but very low accuracy and high log loss. Index (1) and Index (2) are results from \cite{hu2020devicerobust}, where a two-stage system is used. Although they achieve very good performance (80.1\% for two-stage FCNN and 81.9\% for two-stage ensemble), their size is very large, which are 132MB and 332MB, respectively. It should be noted that the reverberation augmented data is used for Index (1) and (2) but not for others.

The Index (3-11) in Table~\ref{tab:task1a} are results of SIC. We here perform the ablation study for each method we propose. Index (3) is the SIC baseline, which has the size of 503KB, accuracy of 67.8\%, and log loss of 0.954. With the use of TSL, shown as Index (4), we can improve the accuracy and log loss while keeping the model size unchanged. We use the two-stage FCNN model, shown as Index (1), as the teacher model. Index (5) shows the result of using LTH, where we can significantly reduce the model size (around $149\times$ compression rate. Although model parameters are reduced in a huge scale, the model performance shows much better than the SIC baseline: Index (3). This verifies our argument that the models are overparameterized a lot.

Index (6) and (7) shows the results by only using two-stage fusion or MTL. From the results, we can see the two-stage can boost the performance, but the method will double the model size. By using a compromise method, MTL, can work in the same manner but save parameters. However, it's slightly worse than using two-stage. Index (8) shows the result by only using quantization. The model parameters are quantized from float32 to float8. Although it obtains a $4\times$ compression rate, the performance worsens when compared with the SIC baseline. However, according to our experiments, we find that the ensemble of 4 quantized models shows better results than an unquantized model, which shows the potential of quantization. With the combination of proposed approaches, we can further boost the performance of SIC model, as shown in Index (9) to (11) of Table~\ref{tab:task1a}. We can at most compress the SIC model to 0.9KB, shown as Index (11), with even better performance than SIC baseline. As for LIC models, shown in Index (12) to (21), the same conclusions as SIC can be observed. Furthermore, when training by augmented data, system robustness can be further boosted. As for LIC, we can at most compress it to 687KB, with the best log loss of 0.925. We can observed from Index (14) and (15) that although the model size decreases (from 60\% to 20\%), we can achieve an even more better accuracy (from 67.64\% to 68.59\%) and log loss.

\textbf{For our final submitted four systems:} four ``two-stage ensembles'' of different LIC and SIC models with LTH pruning are selected. We obtain SICs and LICs from different training epochs by training with different combinations of data augmentation strategies and training criterion (one-hot labels or TS learning). Specifically, for system \textbf{(a)}, we use two 3-class LICs, three 10-class LICs and one 10-class SIC. 
The results of system (1) on development set is specified in Index (22) of Table~\ref{tab:task1a}. The other three submitted systems are not tested on development set due to the time and resources limitation.

\vspace{-0.3cm}

\section{Discussion \& Conclusion}
\label{sec:con}
\vspace{-0.1cm}

As low-complexity acoustic modeling, a lottery ticket hypothesis framework, Acoustic Lottery, is proposed and provides competitive results. As the very first attempt on applying LTH for acoustic learning and modeling, our future works included theoretical analysis on the success of LTH and its relationship between knowledge distillation for different acoustic and robust speech processing tasks~\cite{yang2020characterizing}. We will open source our proposed framework to the community.

\clearpage
\bibliographystyle{IEEEtran}
\bibliography{refs}

\begin{thebibliography}{10}
\providecommand{\url}[1]{#1}
\def\UrlFont{\rmfamily}
\providecommand{\newblock}{\relax}
\providecommand{\bibinfo}[2]{#2}
\providecommand\BIBentrySTDinterwordspacing{\spaceskip=0pt\relax}
\providecommand\BIBentryALTinterwordstretchfactor{4}
\providecommand\BIBentryALTinterwordspacing{\spaceskip=\fontdimen2\font plus
\BIBentryALTinterwordstretchfactor\fontdimen3\font minus
  \fontdimen4\font\relax}
\providecommand\BIBforeignlanguage[2]{{%
\expandafter\ifx\csname l@#1\endcsname\relax
\typeout{** WARNING: IEEEtran.bst: No hyphenation pattern has been}%
\typeout{** loaded for the language `#1'. Using the pattern for}%
\typeout{** the default language instead.}%
\else
\language=\csname l@#1\endcsname
\fi
#2}}

\bibitem{yang2021unsupervised}
D.~Yang, H.~Wang, and Y.~Zou, ``Unsupervised multi-target domain adaptation for
  acoustic scene classification,'' \emph{arXiv preprint arXiv:2105.10340},
  2021.

\bibitem{dcase2016}
A.~Mesaros, T.~Heittola, E.~Benetos, P.~Foster, M.~Lagrange, T.~Virtanen, and
  M.~D. Plumbley, ``Detection and classification of acoustic scenes and events:
  Outcome of the {DCASE} 2016 challenge,'' \emph{IEEE/ACM Transactions on
  Audio, Speech, and Language Processing}, vol.~26, no.~2, pp. 379--393, 2018.

\bibitem{dcase2018}
A.~Mesaros, T.~Heittola, and T.~Virtanen, ``A multi-device dataset for urban
  acoustic scene classification,'' in \emph{Proceedings of the Detection and
  Classification of Acoustic Scenes and Events 2018 Workshop (DCASE2018)},
  November 2018, pp. 9--13.

\bibitem{dcase2017}
A.~Mesaros, T.~Heittola, A.~Diment, B.~Elizalde, A.~Shah, E.~Vincent, B.~Raj,
  and T.~Virtanen, ``{DCASE} 2017 challenge setup: Tasks, datasets and baseline
  system,'' in \emph{Proceedings of the Detection and Classification of
  Acoustic Scenes and Events 2017 Workshop (DCASE2017)}, November 2017, pp.
  85--92.

\bibitem{dcase2020}
\BIBentryALTinterwordspacing
T.~Heittola, A.~Mesaros, and T.~Virtanen, ``Acoustic scene classification in
  dcase 2020 challenge: generalization across devices and low complexity
  solutions,'' in \emph{Proceedings of the Detection and Classification of
  Acoustic Scenes and Events 2020 Workshop (DCASE2020)}, 2020, submitted.
  [Online]. Available: \url{https://arxiv.org/abs/2005.14623}
\BIBentrySTDinterwordspacing

\bibitem{martin2021low}
I.~Mart{\'\i}n-Morat{\'o}, T.~Heittola, A.~Mesaros, and T.~Virtanen,
  ``Low-complexity acoustic scene classification for multi-device audio:
  analysis of dcase 2021 challenge systems,'' \emph{arXiv preprint
  arXiv:2105.13734}, 2021.

\bibitem{hu2020devicerobust}
H.~Hu, C.-H.~H. Yang, X.~Xia, X.~Bai, X.~Tang, Y.~Wang, S.~Niu, L.~Chai, J.~Li,
  H.~Zhu, F.~Bao, Y.~Zhao, S.~M. Siniscalchi, Y.~Wang, J.~Du, and C.-H. Lee,
  ``Device-robust acoustic scene classification based on two-stage
  categorization and data augmentation,'' 2020.

\bibitem{shim2021attentive}
H.-j. Shim, J.-h. Kim, J.-w. Jung, and H.-J. Yu, ``Attentive max feature map
  for acoustic scene classification with joint learning considering the
  abstraction of classes,'' \emph{arXiv preprint arXiv:2104.07213}, 2021.

\bibitem{heittola2020acoustic}
T.~Heittola, A.~Mesaros, and T.~Virtanen, ``Acoustic scene classification in
  dcase 2020 challenge: generalization across devices and low complexity
  solutions,'' \emph{arXiv preprint arXiv:2005.14623}, 2020.

\bibitem{frankle2018lottery}
J.~Frankle and M.~Carbin, ``The lottery ticket hypothesis: Finding sparse,
  trainable neural networks,'' in \emph{International Conference on Learning
  Representations}, 2018.

\bibitem{hu2020two}
H.~Hu, C.-H.~H. Yang, X.~Xia, X.~Bai, X.~Tang, Y.~Wang, S.~Niu, L.~Chai, J.~Li,
  H.~Zhu, \emph{et~al.}, ``A two-stage approach to device-robust acoustic scene
  classification,'' \emph{arXiv preprint arXiv:2011.01447}, 2020.

\bibitem{Chen2019}
H.~Chen, Z.~Liu, Z.~Liu, P.~Zhang, and Y.~Yan, ``Integrating the data
  augmentation scheme with various classifiers for acoustic scene modeling,''
  DCASE2019 Challenge, Tech. Rep., June 2019.

\bibitem{Koutini2019}
K.~Koutini, H.~Eghbal-zadeh, and G.~Widmer, ``Acoustic scene classification and
  audio tagging with receptive-field-regularized {CNNs},'' DCASE2019 Challenge,
  Tech. Rep., June 2019.

\bibitem{mixup}
H.~Zhang, M.~Cisse, Y.~N. Dauphin, and D.~Lopez-Paz, ``mixup: Beyond empirical
  risk minimization,'' \emph{arXiv preprint arXiv:1710.09412}, 2017.

\bibitem{spec-aug}
D.~S. Park, W.~Chan, Y.~Zhang, C.-C. Chiu, B.~Zoph, E.~D. Cubuk, and Q.~V. Le,
  ``Specaugment: A simple data augmentation method for automatic speech
  recognition,'' \emph{arXiv preprint arXiv:1904.08779}, 2019.

\bibitem{spec-corr}
T.~Nguyen, F.~Pernkopf, and M.~Kosmider, ``Acoustic scene classification for
  mismatched recording devices using heated-up softmax and spectrum
  correction,'' in \emph{ICASSP 2020-2020 IEEE International Conference on
  Acoustics, Speech and Signal Processing (ICASSP)}.\hskip 1em plus 0.5em minus
  0.4em\relax IEEE, 2020, pp. 126--130.

\bibitem{hinton2015distilling}
G.~Hinton, O.~Vinyals, and J.~Dean, ``Distilling the knowledge in a neural
  network,'' \emph{arXiv preprint arXiv:1503.02531}, 2015.

\bibitem{li2014learning}
J.~Li, R.~Zhao, J.-T. Huang, and Y.~Gong, ``Learning small-size dnn with
  output-distribution-based criteria,'' in \emph{Fifteenth annual conference of
  the international speech communication association}, 2014.

\bibitem{malach2020proving}
E.~Malach, G.~Yehudai, S.~Shalev-Schwartz, and O.~Shamir, ``Proving the lottery
  ticket hypothesis: Pruning is all you need,'' in \emph{International
  Conference on Machine Learning}.\hskip 1em plus 0.5em minus 0.4em\relax PMLR,
  2020, pp. 6682--6691.

\bibitem{zhou2019deconstructing}
H.~Zhou, J.~Lan, R.~Liu, and J.~Yosinski, ``Deconstructing lottery tickets:
  Zeros, signs, and the supermask,'' \emph{arXiv preprint arXiv:1905.01067},
  2019.

\bibitem{yang2020multi}
C.-H.~H. Yang, L.~Liu, A.~Gandhe, Y.~Gu, A.~Raju, D.~Filimonov, and I.~Bulyko,
  ``Multi-task language modeling for improving speech recognition of rare
  words,'' \emph{arXiv preprint arXiv:2011.11715}, 2020.

\bibitem{Martnmorato2021lowcomplexity}
I.~Martín-Morató, T.~Heittola, A.~Mesaros, and T.~Virtanen, ``Low-complexity
  acoustic scene classification for multi-device audio: analysis of dcase 2021
  challenge systems,'' 2021.

\bibitem{szegedy2015going}
C.~Szegedy, W.~Liu, Y.~Jia, P.~Sermanet, S.~Reed, D.~Anguelov, D.~Erhan,
  V.~Vanhoucke, and A.~Rabinovich, ``Going deeper with convolutions,'' in
  \emph{Proceedings of the IEEE conference on computer vision and pattern
  recognition}, 2015, pp. 1--9.

\bibitem{yang2020characterizing}
C.-H. Yang, J.~Qi, P.-Y. Chen, X.~Ma, and C.-H. Lee, ``Characterizing speech
  adversarial examples using self-attention u-net enhancement,'' in
  \emph{ICASSP 2020-2020 IEEE International Conference on Acoustics, Speech and
  Signal Processing (ICASSP)}.\hskip 1em plus 0.5em minus 0.4em\relax IEEE,
  2020, pp. 3107--3111.

\end{thebibliography}

\end{sloppy}
\end{document}